\documentclass{ECOC-author}
\usepackage{graphicx}
\usepackage{tikz}
\usepackage{lipsum,adjustbox}
\usetikzlibrary{decorations}
\usepackage{pgfplots,siunitx}
\usetikzlibrary{decorations.pathreplacing}
\usetikzlibrary{shapes,arrows,shadows}
\usetikzlibrary{calc,positioning}

\usepackage{subfig}
\usetikzlibrary{external}
\tikzsetfigurename{tikz_compiled} 

\begin{document}

\title{ENUMERATIVE SPHERE SHAPING FOR RATE ADAPTATION AND REACH INCREASE IN WDM TRANSMISSION SYSTEMS}

\author{ Abdelkerim Amari\ad{1}\corr, Sebastiaan Goossens\ad{1}, Yunus Can G\"ultekin\ad{1}, Olga Vassilieva\ad{2}, Inwoong Kim\ad{2}, Tadashi Ikeuchi\ad{2},  Chigo Okonkwo\ad{1},  Frans M J Willems\ad{1}, and Alex Alvarado\ad{1}}

\address{\add{1}{Department of Electrical Engineering, Eindhoven University of Technology, The Netherlands}
\add{2}{Fujitsu Laboratories of America, Inc., 2801 Telecom Parkway, Richardson TX 75082, USA}
\email{a.amari@tue.nl}}


\begin{abstract}
The performance of enumerative sphere shaping (ESS), constant composition distribution matching (CCDM), and uniform signalling are compared at the same forward error correction rate. ESS is shown to offer a reach increase of approximately $10\%$ and $22\%$ compared to CCDM and uniform signalling, respectively. 
\end{abstract}

\maketitle

\section{Introduction}

Probabilistic shaping (PS) has received considerable attention as a promising approach to increase the capacity of optical communication systems \cite{ps0}--\cite{ps6}. In a linear AWGN channel, PS increases the rate by targeting a Gaussian distribution on the transmitted symbols. These gains can be up to $ 0.255$~bit/real dimension (1.53~dB signal-to-noise ratio (SNR) gain) when the number of constellation symbols tends to infinity.

Probabilistic amplitude shaping (PAS) using constant composition distribution matching (CCDM) \cite{ps0} has been widely investigated in fibre optical systems. CCDM has been shown to provide a performance improvement in comparison with uniform signalling. Furthermore, CCDM-based PAS has been proposed for rate adaptation \cite{ps3,ps4}. In this case, the information rate can be adapted by adjusting the distribution of the DM. It has been shown in \cite{ps4} that PS-$64$-QAM provides better performance than uniform $16$-QAM, operating at the same forward error correction (FEC) and information rate.


CCDM-based PAS suffers from significant rate losses at short blocklengths. For optimum performance, long blocklengths are therefore required. In optical communication systems, this causes a nonlinear penalty \cite{ps2}. Additionally, long-blocklengths CCDM is very challenging to implement in real-time. To palliate the rate loss problem, multiset-partition DM has been proposed in \cite{mp}. 

We recently proposed in \cite{amari2019} to use PAS in combination with enumerative sphere shaping (ESS) \cite{gultekin1,gultekin2} as an alternative way to reduce the rate loss whilst using short blocklengths. In the context of wavelength-division multiplexing (WDM) systems, it was shown in \cite{amari2019} that ESS provides linear shaping gains in comparison with CCDM due its low rate loss. The comparison between PAS and uniform signalling in \cite{amari2019} was done for the same modulation format but different FEC rates. Furthermore, short blocklength ESS and CCDM were shown in \cite{amari2019} to provide higher nonlinear tolerance than PS with long blocklengths, and also higher than uniform signalling. The results in \cite{amari2019} show that short blocklength ESS-based PAS is a promising shaping approach because it combines both linear shaping gain and nonlinear tolerance.

In this paper, we propose to use ESS for rate adaptation and study its potential gains. ESS-based PS-$256$-QAM is compared against CCDM-based PS-$256$-QAM as well as to uniform $64$-QAM \emph{at the same FEC rate} and information rate. The rate adaptation for ESS is performed by adjusting the maximum energy, which is a tunable parameter in ESS-based PAS \cite{gultekin1,amari2019}. For a $450$ Gb/s dual-polarization WDM system, ESS with blocklengths of $200$ and $100$ increase the transmission reach by approximately $10\%$ and $22\%$ when compared to CCDM with $200$ and uniform.

\section{System Model}\label{model}
\begin{figure*}[h]
	\centering		
	\includegraphics[]{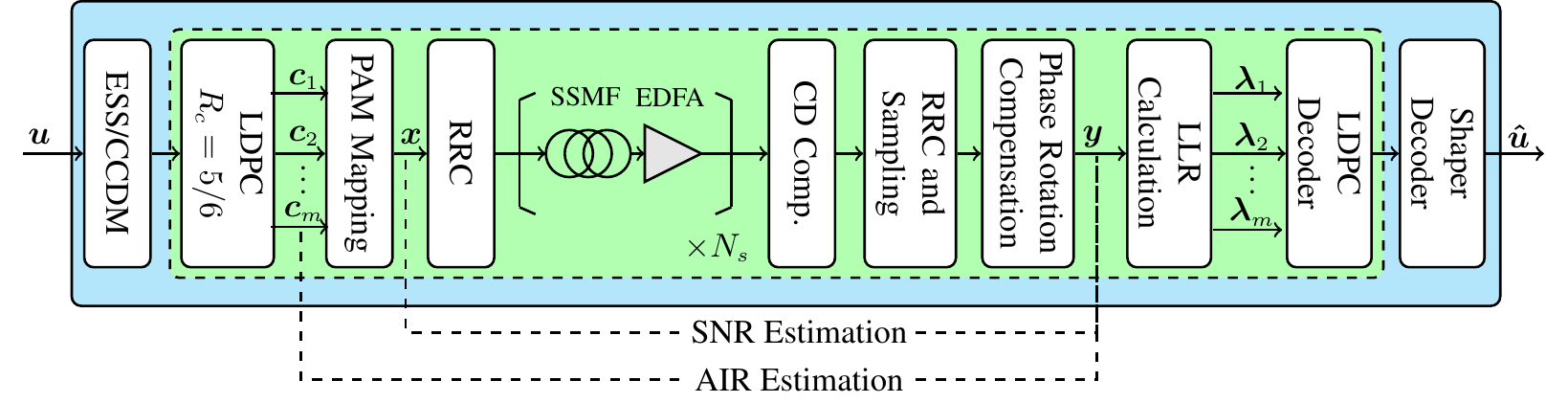}
		\vspace{-0.2cm}
		\caption{Simulation setup with a fixed LDPC rate. Uniform signalling with $m=3$ (green). PAS with $m=4$ (blue).
		}	
		    \label{fig:1}
\end{figure*}

We compare PS-$256$-QAM based on ESS and CCDM with uniform $64$-QAM. The main idea is to adjust the shaping rate of ESS/CCDM to have the same information rate as uniform $64$-QAM. We consider $M$-QAM signals obtained by the Cartesian product of two identical $2^m$-ary pulse amplitude modulation (PAM) constellations using the binary reflected Gray code as labelling. Without loss of generality, we focus on the $2m$-PAM constellation. We use a low-density parity-check (LDPC) coding  with a rate $R_c=5/6$. The information rate of  uniform $8$-PAM is $R = m R_c= 2.5$~[bits/1D-sym], where $m=3$ is the the number of bit-levels. The transmitter side in Fig.~\ref{fig:1} shows this structure.

For the shaping schemes, we consider $256$-QAM, i.e., $m=4$. In this case, the information rate is adapted by adjusting the shaping rate. The shaping rate of PAS is defined as  $R_s = k/N$~[bits/amp] where $k$ is the number of input bits and $N$ is the number of output amplitudes of the shaper. While CCDM fixes the probability distribution and generates the sequences that satisfy this distribution, ESS uses bounded energy sequences by putting a maximum energy constraint \cite{gultekin1,amari2019}.
Therefore, for CCDM, the shaping rate is adjusted by changing the input distribution of the shaper. For ESS, the shaping rate is adjusted by changing the maximum energy.

The information rate per real dimension ($1$D-sym) for PAS is given by $R=(k+\gamma N)/N=R_s + \gamma = 2.5$~[bits/1D-sym], where $0 < \gamma < 1$ represents additional information bits. The value of $\gamma$ is determined from the LDPC coding rate $R_c$ as $\gamma =m(R_{c} - 1) + 1 =0.33$ ~[bits/1D-sym], where we used $m=4$($16$-PAM). Thus, the shaping rate is given by $R_s= R - \gamma = 2.16$~[bits/amp]. Note that the outputs of the shaper are one-sided positive amplitudes, and the sign bits are added at the FEC coding level via the PAS framework.

\section{Simulation Setup and Results}

We consider a dual-polarization long-haul WDM transmission system with a net bit rate per WDM channel of \mbox{$450$ Gb/s}. This net bit rate is obtained by considering an LDPC code with $R_c=5/6$, a symbol rate per polarization of $45$ Gbaud, and an information rate of $2.5$~[bits/1D-sym] as described in Sec.~\ref{model}. $11$ WDM channels are transmitted with a channel spacing of $50$ GHz. For the transmission link, a multi-span standard single mode fibre (SSMF) is used with attenuation $\alpha=0.2~\mathrm{dB\cdot km^{-1}}$,  dispersion parameter $D=17 ~\mathrm{ps \cdot nm^{-1} \cdot km^{-1}}$, and nonlinear coefficient \mbox{$\Gamma=1.3~ \mathrm{W^{-1} \cdot km^{-1}}$}. The signal amplification at each span of $L=80$~km is ensured using an erbium-doped fibre amplifier (EDFA) with a $5$ dB noise figure and $16$ dB gain.

Fig.~\ref{fig:1} shows the simulation setup for each real-dimension of the transmitted signal. The uniform signalling scheme is represented by the dashed green box, and the shaping scheme is represented by the outer blue box.
At the transmitter side, the information bits $\boldsymbol{u}=(u_1, u_2, \ldots, u_k)$  are shaped using ESS or CCDM. After the LDPC coding, the coded bits $\boldsymbol{c} = (\boldsymbol{c}_1, \boldsymbol{c}_2, \ldots,\boldsymbol{c}_m)$  are mapped to PAM symbols $\boldsymbol{x}$. Then, the signal is oversampled with $16$ samples/symbol. A root-raised cosine (RRC) filter, with roll-off factor of $0.1$, is used for spectrum shaping.

At the receiver side, channel selection is firstly applied and then, the signal is downsampled to $2$ samples/symbol. After that chromatic dispersion (CD) compensation is performed before applying an RRC matched filter and downsampling to $1$ sample/symbol. An ideal phase rotation compensation is performed. Then, log-likelihood ratios (LLRs) $\boldsymbol{\lambda}_1, \boldsymbol{\lambda}_2, \ldots,\boldsymbol{\lambda}_m$ are calculated and passed to the soft-decision LDPC decoder. Finally, ESS or CCDM deshaping is performed to recover the transmitted information bits.

 The results will be given in terms of finite blocklength bit-metric decoding (BMD) rate, which gives an indication of the overall performance, and the effective SNR, which is a good metric to evaluate the nonlinear tolerance. The finite blocklength BMD rate is defined as (see~\cite[eq.~(26)]{mp})
 \begin{equation}\label{gmi}
     \text{AIR}_{\text{N}} = \underbrace{\left[  H(\boldsymbol{C}) - \sum_{i=1}^m H(C_i \mid Y) \right]}_{\text{BMD Rate}} - \underbrace{\left[H(A)-\frac{k}{N}\right]}_{\text{Rate loss}},
 \end{equation}
where $H(\cdot)$ denotes entropy and $A$ are the shaped amplitudes.

 The effective SNR is defined as \cite{ps3} 
\begin{align}\label{eff.snr}
\text{SNR}_{\text{eff}} \approx \frac{\mathbb{E}[|{X}|^2]}{\mathbb{E}[|{Y}-{X}|^2]}, 
\end{align}
where $\mathbb{E}[\cdot]$ represents expectation and $X$ and $Y$ are the transmitted and received symbols respectively. The effective SNR is calculated per QAM symbol and takes into account the probability of each constellation points.


 We focus on evaluating the performance of the center WDM channel because it is more affected by the channel impairments. We compare five different schemes: ESS and CCDM at the same blocklength $N=200$, long blocklength CCDM ($N=3600$), uniform signalling, and ESS with ultra short blocklength ($N=100$).

\begin{figure}[t]
	\centering		
		\includegraphics[width=0.93\linewidth]{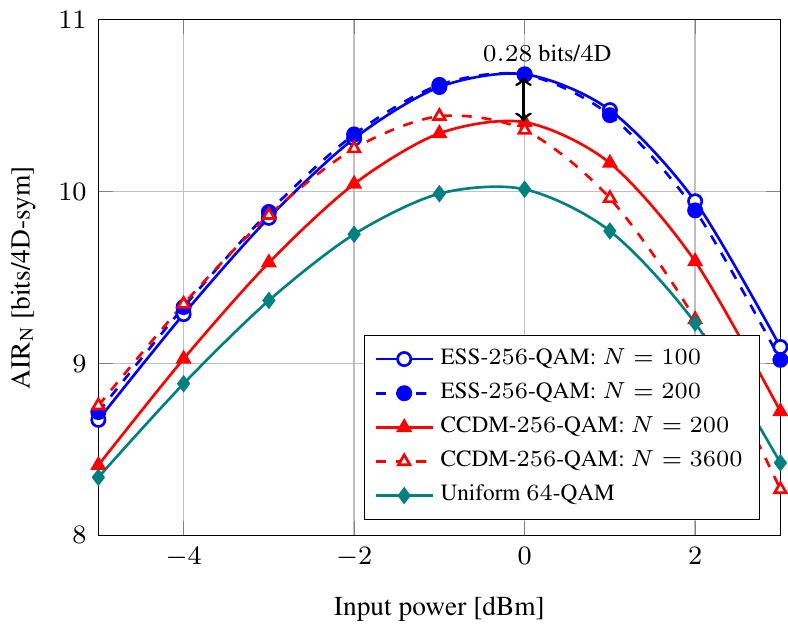}
		\caption{$\text{AIR}_{\text{N}}$ vs. input power: PS-$256$-QAM and uniform $64$-QAM at $1600$ km.}	
		\label{fig:2}
\end{figure}

In Fig.~\ref{fig:2}, we plot the finite blocklength BMD rate in \eqref{gmi} versus the input power for a transmission distance of $1600$ km. It is observed that, in the linear regime at low input powers, CCDM with $N=3600$ exhibits slightly higher performance than short blocklengths ESS, which outperforms short blocklength CCDM at $N=200$. The linear regime results are explained by the rate loss \cite{gultekin1,amari2019}, in which ESS outperforms CCDM at short blocklengths.  Long blocklength CCDM ($N=3600$) is better than ESS with $N=200$ and $N=100$ because the shorter the blocklength, the higher the rate loss.
At optimal input power ($0$~dBm), ESS with $N=100$ and $N=200$ have similar finite blocklength performance and show a gain of about $0.28$ bits/$4$D-sym in comparison with CCDM at $N=200$. Gains of $0.33$ bits/$4$D-sym and $0.68$ bits/$4$D-sym with respect to CCDM at $N=3600$ and uniform signalling, respectively, are also shown.
It is observed that the optimal power of short blocklength ESS and CCDM ($0$ dBm) is higher than the optimal power of long blocklength CCDM ($-1$ dBm). Furthermore, in the nonlinear regime (above the optimum launch power), short blocklength ESS and CCDM show higher finite blocklength BMD than CCDM with $N=3600$. This will be explained by effective SNR results, which evaluate the impact of the fibre nonlinearity.

\begin{figure}[t]
	\centering		
		\includegraphics[width=0.93\linewidth]{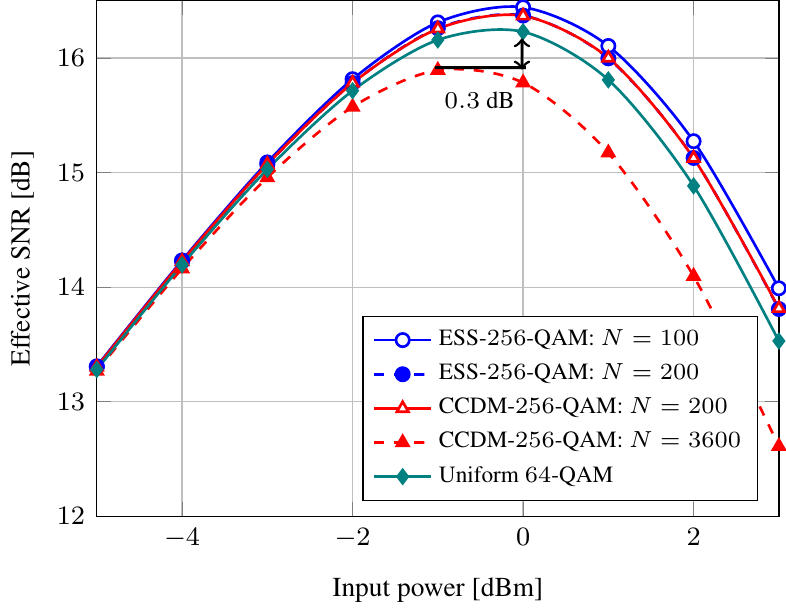}
		\caption{Effective SNR vs. input power: PS-$256$-QAM and uniform $64$-QAM at $1600$ km.}	
		\label{fig:3}
\end{figure}
The effective SNR in \eqref{eff.snr} is plotted as a function of the input power in Fig.~\ref{fig:3} at the same transmission distance of Fig.~\ref{fig:2} ($1600$ km). Firstly, we observe that uniform signalling shows higher effective SNR than CCDM with $N=3600$. The gain is about $0.3$ dB. This coincides with state of the art results in which long blocklength CCDM is less tolerant to the fibre nonlinearity than uniform \cite{ps2}. We also observe that ESS and CCDM exhibits similar performance in terms of effective SNR at the same block length $(N=200)$. ESS with a shorter blockength ($N=100$) gives a slightly larger SNR, but at the same time, has a larger rate loss than ESS with $N=200$.
The results in Fig.~\ref{fig:3} show that the impact of the fibre nonlinearities depends on the blocklength, and that long blocklength shaping should be avoided in the nonlinear fibre channel. 

\begin{figure}[t]
	\centering		
		\includegraphics[width=0.93\linewidth]{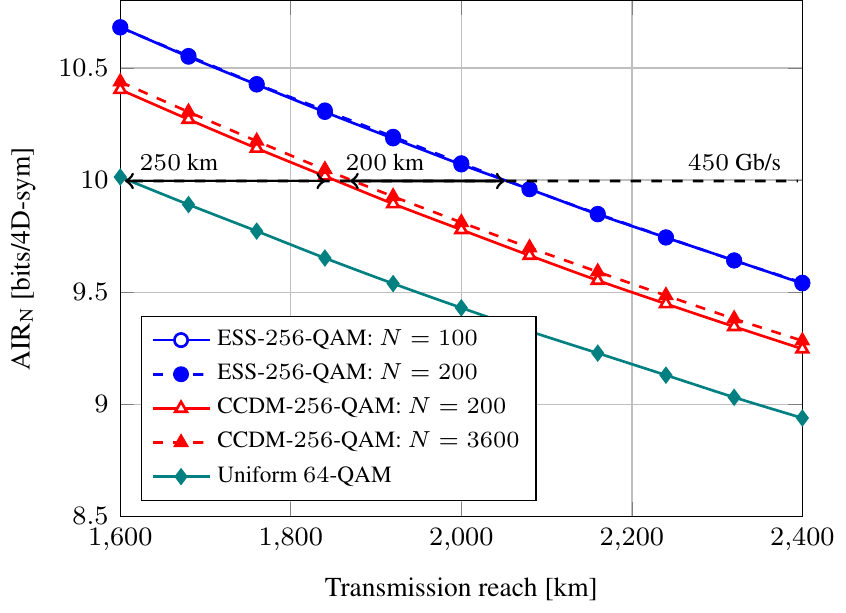}
		\caption{$\text{AIR}_{\text{N}}$vs. transmission reach: PS-$256$-QAM and uniform $64$-QAM at optimal input power.}	
		\label{fig:4}
\end{figure}
In Fig.~\ref{fig:4}, we plot the finite blocklength BMD versus transmission reach at optimal input powers. It is observed that ESS with $N=100$ and $N=200$ exhibits similar performance in terms of transmission reach. Even though ESS with $N=200$ has lower rate loss than $N=100$, the rate loss difference is compensated by the nonlinear tolerance gain that ESS with $N=100$ provides (see Fig.~\ref{fig:3}). At the net bit rate of $450$ Gb/s, which corresponds to an $\text{AIR}_{\text{N}}=10$ bits/$4$D, short blocklengths ESS outperforms CCDM and uniform signalling by about $200$ km  and $450$ km, respectively.

\section{Conclusions}

We have proposed to use enumerative sphere shaping for rate adaptation and performance improvement in the context of $450$ Gb/s dual-polarization WDM system. We have shown that rate adapted- PS-$256$-QAM increases the performance in comparison with uniform $64$-QAM, and that short blocklength ESS-based $256$-QAM outperforms CCDM-based $256$-QAM with short and long blocklengths. Furthermore, short blocklength shaping provides higher nonlinear tolerance and lower complexity than long blocklength case, which make it more suitable to be used for optical communication systems. Due to it low rate loss, nonlinear tolerance, and lower complexity, short blocklengths ESS is shown to be a better alternative to CCDM to be implemented in real-time high-speed systems.

\vspace{0.7ex}
{\small
\noindent\textbf{Acknowledgements:}
This work was supported by the Netherlands Organization for Scientific Research (NWO) via the VIDI Grant ICONIC (project number 15685). The work of A. Alvarado has received funding from the European Research Council (ERC) under the European Union's Horizon 2020 research and innovation programme (grant agreement No 757791). 
}

\end{document}